\documentclass[12pt,a4paper]{article}
\usepackage{amsmath,amsthm,amsfonts,amssymb,alltt}
\usepackage{graphics}
\usepackage{graphicx}
\usepackage{epsfig}

\newtheorem{theorem}{Theorem}[section]
\newtheorem{lemma}[theorem]{Lemma}

\theoremstyle{definition}

\newtheorem{example}[theorem]{Example}
\theoremstyle{remark}
\newtheorem{remark}[theorem]{Remark}

\numberwithin{equation}{section}

\begin{document}

\author{Vladimir Mityushev \\
Dept. Computer Sciences and Computer Methods, \\ Pedagogical University,\\
Podchorazych 2, Krakow 30-084, Poland}
\title{Pattern formations and optimal packing}
\date{}
\maketitle

\begin{abstract}
Patterns of different symmetries may arise after solution to reaction-diffusion equations. Hexagonal arrays, layers and their perturbations are observed in different models after numerical solution to the corresponding initial-boundary value problems. We demonstrate an intimate connection between pattern formations and optimal random packing on the plane. The main study is based on the following two points. First, the diffusive flux in reaction-diffusion systems is approximated by piecewise linear functions in the framework of structural approximations. This leads to a discrete network approximation of the considered continuous problem. Second, the discrete energy minimization yields optimal random packing of the domains (disks) in the representative cell. 
Therefore, the general problem of pattern formations based on the reaction-diffusion equations is reduced to the geometric problem of random packing. It is demonstrated that all random packings can be divided onto classes associated with classes of isomorphic graphs obtained form the Delaunay triangulation. The unique optimal solution is constructed in each class of the random packings. If the number of disks per representative cell is finite, the number of classes of isomorphic graphs, hence, the number of optimal packings is also finite.  
\end{abstract}

\section{Introduction}
The Turing mechanism for reaction-diffusion equations models biological and chemical pattern formations. This approach was widely discussed in literature and supported by many numerical examples (see the recent books \cite{Bressloff}, \cite{Jost}, \cite{Walf} and many works cited therein). 
Patterns of different symmetries may arise after solution to reaction-diffusion equations. Hexagonal arrays, layers and their perturbations are observed in different models after numerical solution to the corresponding initial-boundary value problems for nonlinear partial differential equations. 
However, these models do not answer the question, why the most frequently observed patterns are close to the optimal packing structures. Why do the  hexagonal array arise? One can see, for instance, that a resulting structure can be the hexagonal array disturbed by pentagon inclusions. Is it related to a model approximation or to an inherent feature of pattern formations?  

In the present paper, we try to answer the above questions to demonstrate an intimate connection between pattern formations and optimal random packing on the plane. The main study is based on the following two points. First, the diffusive flux in reaction-diffusion systems is approximated by piecewise linear functions in the framework of structural approximations \cite{LB}, \cite{Kolpakov}. This leads to a discrete network approximation of the considered continuous problem. Second, the discrete energy minimization yields optimal random packing of the domains in the representative cell. The packed domains are approximated by equal disks. These approach is described in the bulk of the paper.     

Packing problems refers to geometrical optimization problems \cite{Toth}. In the present paper, we consider the optimal packing of disks on the plane in the random statement fitted to the description of pattern formations. 
Optimal packing in the classic deterministic statement is attained for the hexagonal array when the packing concentration holds $\frac{\pi}{\sqrt{12}}$ \cite{Toth}.  Computer simulations demonstrate that random packing have a lower density and depends on the protocol of the random packing. 

It is shown in Sec.\ref{secM33} that pattern formations lead to the optimal random packing problem. In the present paper, this problem is resolved by introduction of the equivalence classes of graphs obtained form the Delaunay triangulation associated to packings. The justification of such an approach is based on the observation  that solution to the physical problem of the optimal diffusion implies solution to the geometrical problem of the packing disks \cite{Mit2012}. The unique optimal solution is constructed in each class of the random packings. If the number of disks per representative cell is finite, the number of classes of isomorphic graphs, hence, the number of optimal packings is also finite. 
 
The proposed method to study pattern formations is based on the investigation of graph structures by analytical and numerical methods without treatment of PDE.  

\section{Structural approximation}
\label{sec2}
\begin{figure}[htp]
\centering
\includegraphics[clip, trim=0mm 0mm 0mm 0mm, width=1.0\textwidth]{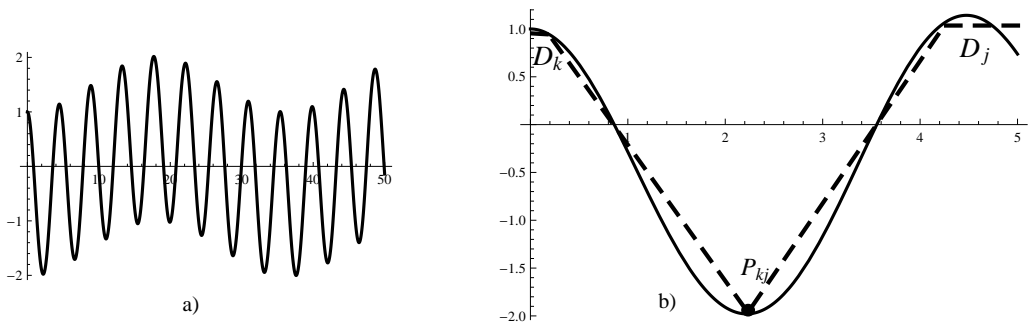} 
\caption{a) Dependence of the inhibitor on the spatial variable. b) Piecewise linear approximation of the inhibitor on a smaller interval (dashed line). The maxima are approximated by segments $D_k$ and $D_j$ (disks in 2D) and the minima by points $P_{kj}$ (segments in 2D).} 
\label{fig:bio1}
\end{figure}

The Turing mechanism can create temporally stable and spatially non-homogeneous structures. In order to present the main idea of the structural approximation we consider 1D Schnakenberg system \cite[p. 156]{Jost}. A typical dependence of the inhibitor on the spatial variable is displayed in Fig.\ref{fig:bio1}a. It is assumed that such a dependence can be approximated by a  piecewise linear function as shown in Fig.\ref{fig:bio1}b.   
The solution of the continuous reaction-diffusion equations is approximated by the discrete diffusion model with the constant diffusion fluxes (derivatives of the linear approximations) between the extrema of the potential.
 
\begin{figure}[htp]
\centering
\includegraphics[clip, trim=0mm 0mm 0mm 0mm, width=0.8\textwidth]{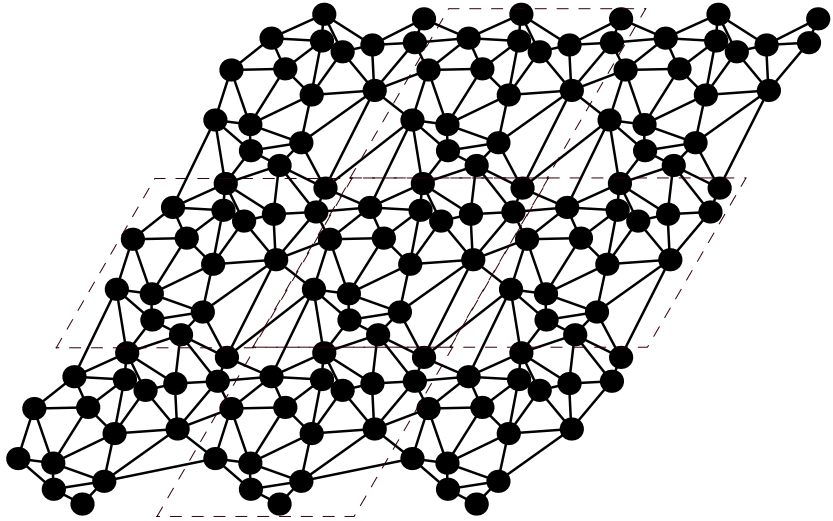} 
\caption{2D approximation of the inhibitor. The diffusion potential is approximated by appropriate constants in disks and the diffusion flux between the disks by linear functions along the edges of the Delaunay triangulation.} 
\label{fig:bio4}
\end{figure}
A similar  approximation can be extended to multidimensional reaction-diffusion equations \cite{Mit2014}. In the present paper, we deal with 2D double periodic structures. Let $\mathbf e_1=(e_1,0)$ and $\mathbf e_2=(e_{21},e_{22})$ be the translation vectors of the lattice $\mathcal Q =\{l_1\mathbf e_1+l_2\mathbf e_2: \; l_j \in \mathbb Z \}$ where $\mathbb Z$ denotes the set of integer numbers. 
Consider the periodic representative cell 
$$
Q_{0} = \{\mathbf x =t_1 \mathbf e_1 +t_2 \mathbf e_2\}, \;0 < t_j <1 \}.
$$ 
For simplicity, we approximate the places of maximal diffusion potential by equal disks $D_i$ ($i=1,2,\ldots,N$) of radius $r$ centered at the set of points $\mathbf A=(\mathbf a_1,\mathbf a_2, \ldots, \mathbf a_N)$  displayed in Fig.\ref{fig:bio4}. The maxima of the diffusion potential are approximated by disks and the minima by segments\footnote{It is natural to introduce the disk approximations also for minima. But in this case we shall obtain two types of disks that complicates the one-type-disks model constructed below. Because the diffusion flux will occur between the disks of different types and vanish between the disks of the same type.}. Every line segment $P_{kj}$ is perpendicular to the segment $(\mathbf a_k,\mathbf a_j)$, its length holds $|P_{kj}|=2r$ and it is divided onto equal parts by $(\mathbf a_k,\mathbf a_j)$. 

It is convenient to introduce new distance (metric) as follows. Two points $\mathbf a, \mathbf b \in \mathbb R^2$ are identified if their difference $\mathbf a- \mathbf b=l_1\mathbf e_1+l_2\mathbf e_2$ belongs to the lattice $\mathcal Q$. Hence, the classic flat torus topology with the opposite sides welded  is introduced on the cell $Q_{0}$. The distance $\|\mathbf a-\mathbf b\|$ between two points $\mathbf a,\mathbf b \in Q_0$ is introduced as
\begin{equation}\|\mathbf a-\mathbf b\|:=\min_{l_1,l_2 \in \mathbb Z} \left|\mathbf a-\mathbf b+l_1\mathbf e_1+l_2\mathbf e_2\right|,
\label{eq:sam1}
\end{equation} 
where the modulus means the Euclidean distance  in $\mathbb R^2$ between the points $\mathbf a$ and $\mathbf b$.

Construct the double periodic Voronoi diagram and the Delaunay triangulation corresponding to the set $\mathbf A$ on the torus $\mathcal Q_0=\cup_{l_1,l_2  \in \mathbb Z}(Q_0+l_1\mathbf e_1+l_2\mathbf e_2)$. The edges of the Delaunay triangulation $E$ correspond to linear approximations of the diffusion flux between disks. The Delauney triangulation of the vertices $\mathbf A$ consists of straight lines connecting by pairs points of $\mathbf A$ belonging to neighbor Voronoi regions\footnote{The terms the Delaunay triangulation and graph used in this paper are slightly different from the commonly used notations in degenerate cases. For example, consider a square and its four vertices. The traditional Delaunay triangulation has four sides of the square and one of the diagonals. In our approach, the Delaunay graph has only four sides.}. Let the neighborhood relation between two vertexes be denoted by $\mathbf a_j \sim \mathbf a_k$ or shortly $j \sim k$. We call the constructed double periodic graph $(\mathbf A,E)$ by the Delaunay graph. 

Two graphs are called isomorphic if they contain the same number of vertices connected in the same way. One of the most important notation of the persent paper is the class of graphs $\mathcal G=\mathcal G_{(\mathbf A,E)}$ isomorphic to the given graph $(\mathbf A,E)$. 

Let $\mathbf u=(u_1,u_2, \ldots, u_N)$ denote the vector whose components are the maximal diffusion potentials in the corresponding disks.   
The discrete network model for densely packed disks \cite{LB}, \cite{Kolpakov}, \cite{ryl2008} is based on the fact that the diffusion flux is concentrated in the necks between closely spaced inclusions having different potentials. In our model, closely spaced inclusions means the chain disk--segment--disk ($D_k\leftrightsquigarrow P_{kj}\leftrightsquigarrow D_j$) displayed in Fig.\ref{fig:bio1}b. 
For two neighbor disks $D_k$, $D_j$ with centers $\mathbf a_k$, $\mathbf a_j$ and a segment $P_{kj}$ between them the relative interparticle flux $g(|\mathbf a_k - \mathbf a_j|)$ can be approximated by Keller's formula \cite{Keller} 
\begin{equation}
g(|\mathbf a_k - \mathbf a_j|)=\pi \sqrt{\frac{r}{\delta_{kj}}},
\label{eq:keller2}
\end{equation}
where $\delta_{kj} =\|\mathbf a_k - \mathbf a_j\| - 2r$ denotes the gap between the neighbor disks. Keller's formula \eqref{eq:keller2} was deduced for the linear local diffusion flux between the neighbor disks that is agree with our approximation.  
Introduce the designation 
\begin{equation}
 \mathop{{\sum_{k,j}}^{(\mathcal G)}} = \sum_{k=1}^N \sum_{j \sim k},
\label{eq:sa2d}
\end{equation}
where $j \sim k$ means that the vertices $\mathbf a_j$ and $\mathbf a_k$ are connected. Following \cite{LB}, \cite{Kolpakov} introduce the functional associated to the discrete energy  
\begin{equation}
E(\mathbf u, \mathbf a)=\mathop{{\sum_{k,j}}^{(\mathcal G)}} g(\|\mathbf a_k - \mathbf a_j\|) (u_k+u_j)^2,
\label{eq:en1}
\end{equation}
where $u_k+u_j$ is the variation of the diffusion potential along the chain $D_k\leftrightsquigarrow P_{kj}\leftrightsquigarrow D_j$.

\section{Optimal random packing}
\label{secM33}
Consider the minimization problem
\begin{equation}
\mathcal E(\mathbf u)= \min_{\mathbf A}\;E(\mathbf u,\mathbf A) = \min_{\mathbf A}\mathop{{\sum_{k,j}}^{(\mathcal G)}} g(\|\mathbf a_k - \mathbf a_j\|) (u_k+u_j)^2.
\label{eq:sa1f6}
\end{equation} 
The function $g(x)= \pi \sqrt{\frac{r}{x-2r}}$ as a convex function satisfies Jensen's inequality
\begin{equation}
\sum_{i=1}^M p_i g(x_i) \geq g \left(\sum_{i=1}^M p_i x_i \right),
\label{eq:sa5}
\end{equation} 
where the sum of positive numbers $p_i$ is equal to unity. Equality holds if and only if all $x_i$ are equal. Let the sum from \eqref{eq:en1} is arranged in such a way that $x_i=\|\mathbf a_k-\mathbf a_j\|$ and $p_i=\frac 1U(u_k+u_j)^2$, where $U=\mathop{{\sum}^{(\mathcal G)}} (u_k+u_j)^2$. 
Application of \eqref{eq:sa5} to \eqref{eq:en1} yields
\begin{equation}
\mathop{{\sum}^{(\mathcal G)}}  g(\|\mathbf a_k-\mathbf a_j\|)(u_k +u_j)^2
\geq U g \left(\frac 1U \mathop{{\sum}^{(\mathcal G)}}  
(u_k +u_j)^2 \|\mathbf a_k-\mathbf a_j\| \right).
\label{eq:sa6}
\end{equation} 
H\"{o}lder's inequality states that for non-negative $a_i$ and $b_i$
\begin{equation}
\sum_{i=1}^M a_i b_i \leq  \left(\sum_{i=1}^M a_i^2 \right)^{\frac 12}  \left(\sum_{i=1}^M b_i^2 \right)^{\frac 12}.
\label{eq:sa7}
\end{equation} 
This implies that 
\begin{equation}
\mathop{{\sum}^{(\mathcal G)}}   (u_k+u_j)^2 \|\mathbf a_k-\mathbf a_j\| \leq \left[ \mathop{{\sum}^{(\mathcal G)}}  (u_k+u_j)^4 \right]^{\frac 12} \left[\mathop{{\sum}^{(\mathcal G)}}\|\mathbf a_k-\mathbf a_j\|^2 \right]^{\frac 12}.
\label{eq:sa8}
\end{equation} 
The function $g(x)$ decreases, hence \eqref{eq:sa6} and \eqref{eq:sa8} give
\begin{equation}
\frac 1U \mathop{{\sum}^{(\mathcal G)}}  g(\|\mathbf a_k-\mathbf a_j\|)(u_k +u_j)^2
\geq  
g \left(\frac 1U \left[ \mathop{{\sum}^{(\mathcal G)}}  
(u_k+u_j)^4 \right]^{\frac 12} \left[\mathop{{\sum}^{(\mathcal G)}} \|\mathbf a_k-\mathbf a_j\|^2 \right]^{\frac 12} \right).
\label{eq:sa9}
\end{equation} 
The minimum of the right hand part of \eqref{eq:sa9} on $\mathbf A$ is achieved independently on $u_k$ for $\max_{\mathbf A} h(\mathbf A)$ where
\begin{equation}
h(\mathbf A)=\mathop{{\sum}^{(\mathcal G)}}  \|\mathbf a_k-\mathbf a_j\|^2.
\label{eq:sa10}
\end{equation}

\begin{lemma}[\cite{Mit2014}]
\label{lemm1}
For any fixed class $\mathcal G_{(\mathbf A},E)$, every local maximizer of $h(\mathbf A)$ is the global maximizer which fulfils the system of linear algebraic equations
\begin{equation}
\mathbf a_k=\frac 1{N_k} \sum_{j \sim k} \mathbf a_j+\frac 1{N_k} \sum_{\ell=1,2} s_{k\ell}\mathbf e_{\ell},\quad k=1,2,\ldots,N,
\label{eq:sa11}
\end{equation}
where $s_{k\ell}$ can take the values $0,\pm 1, \pm 2$ in accordance with the class $\mathcal G_{(\mathbf A,E)}$. 
The system \eqref{eq:sa11} has always a unique solution up to an additive arbitrary constant vector. 
\end{lemma}

Equations \eqref{eq:sa11} describe the stationary points of the functional \eqref{eq:sa10} obtained by its differentiation on $\mathbf a_k$ ($k=1,2,\ldots,N$) 
\begin{equation}
\sum_{j \sim k} (\mathbf a_j-\mathbf a_k) \equiv \mathbf 0.
\label{eq:sa21}
\end{equation}
Here, the congruence relation $\mathbf a \equiv \mathbf b$ means that $\mathbf a - \mathbf b =l_{1}\mathbf e_1+l_{2}\mathbf e_2$ for some integer $l_{1,2}$. Therefore, a point $\mathbf a$ on the torus $\mathcal Q_0$ is associated to the infinite set of points $\{\mathbf a+l_{1}\mathbf e_1+l_{2}\mathbf e_2, \;l_{1,2}\in \mathbb Z\}$ on the plane $\mathbb R^2$. We now rewrite equation \eqref{eq:sa21} on the torus as an equation on the plane for a fixed point $\mathbf a_k \in Q_0$. Consider a points $\mathbf a'_j \in \mathbb R^2$ neighboring to $\mathbf a_k$, i.e.,  $j \sim k$ in a graph $(\mathbf A,E) \in \mathcal G_{(\mathbf A,E)}$. The point $\mathbf a'_j$ is congruent to a point $\mathbf a_j \in Q_0$. 
The graph $(\mathbf A,E)$ corresponds to the Voronoi tessellation, hence, $\mathbf a'_j$ belongs to $Q_0$ or to neighbor cells $Q_0 \pm \mathbf e_1$, $Q_0 \pm \mathbf e_2$, $Q_0 \pm \mathbf e_1 \pm \mathbf e_2$. Therefore, 
$\mathbf a'_j = \mathbf a_j + l_{1jk}\mathbf e_1+l_{2jk}\mathbf e_2$, where $l_{1jk}$ and $l_{2jk}$ can be equal only to $0, \pm 1$. Then, equations \eqref{eq:sa21} can be written in the form \eqref{eq:sa11}
where 
\begin{equation}
s_{1k} = \sum_{j \sim k} l_{1jk}, \quad s_{2k} = \sum_{j \sim k} l_{2jk}.
\label{eq:sam23}
\end{equation} 

One can see that the sum of all equations \eqref{eq:sa11} gives an identity, hence, they are linearly dependent. Moreover, if $\mathbf A = (\mathbf a_1,\mathbf a_2, \ldots, \mathbf a_N)$ is a solution of \eqref{eq:sa11}, then $(\mathbf a_1+\mathbf c,\mathbf a_2+\mathbf c, \ldots, \mathbf a_N+\mathbf c)$ is also a solution of \eqref{eq:sa11} for any $\mathbf c \in \mathbb R^2$. Let the point $\mathbf a_N$ be arbitrarily fixed. Then, $\mathbf a_1,\mathbf a_2, \ldots, \mathbf a_{N-1}$ can be found from the uniquely solvable system of linear algebraic equations
\begin{equation}
\mathbf a_k=\frac 1{N_k} \sum_{j \sim k} \mathbf a_j+\frac 1{N_k} \sum_{\ell=1,2} s_{k\ell}\mathbf e_{\ell},\quad k=1,2,\ldots,N-1.
\label{eq:sa24}
\end{equation}
It is worth noting that the system \eqref{eq:sa24} can be decomposed onto two independent systems of scalar equations on the first and second coordinates of the points $\mathbf a_1,\mathbf a_2, \ldots, \mathbf a_{N-1}$. 

\section{Conclusion and numerical examples}
\label{secNum}
We now proceed to summarize the algorithm to solve the optimization problem. 
First, let a class of graphs $\mathcal G_{(\mathbf A,E)}$ be fixed with the corresponding translation vectors $\mathbf e_1$ and $\mathbf e_2$. Further, the constants $s_{k\ell}$ are constructed by the scheme described at the end of the previous section. The main numerical step is solution to the uniquely solvable system of linear algebraic equations \eqref{eq:sa24} to construct vertices $\mathbf A$ and the corresponding double periodic Delaunay graph $(\mathbf A,E)$. This  graph is called optimal in the class  $\mathcal G_{(\mathbf A,E)}$. The optimal graph not necessary does correspond to a Voronoi tessellation. In this case, one can change a class of graphs by introduction of the new Voronoi tessellation for the vertices $\mathbf A$. Then, the set $\mathbf A$ will not necessary be optimal in the new class $\mathcal G'_{\mathbf A}$. Let $(\mathbf A',E')$ be the optimal graph in the class $\mathcal G'_{\mathbf A}$. Next, if the graph $(\mathbf A',E')$ does not correspond to a Voronoi tessellation, it can be "improved" by $(\mathbf A'',E'')$ etc. Therefore, we arrive at the graph chain 
\begin{equation}
(\mathbf A,E) \to(\mathbf A',E') \to (\mathbf A'',E'') \to \cdots.
\label{eq:sa25}
\end{equation} 

\begin{example}
\label{ex:1}
Consider the hexagonal lattice defined by the fundamental translation vectors $\mathbf e_1=\sqrt[4]{\frac{4}{3}}(1,0)$ and $\mathbf e_2= \sqrt[4]{\frac{4}{3}}(\frac{1}2,\frac{\sqrt{3}}2)$. The area of the cell $Q_0$ holds unit. Consider $N=3$ points $(1.075, 0.175), (0.919, 0.553), (0.444, 0.169)$ and the corresponding double periodic Voronoi tessellation shown in Fig.\ref{fig:ex1}a.
\begin{figure}[htp]
\centering
\includegraphics[clip, trim=0mm 0mm 0mm 0mm, width=1.0\textwidth]{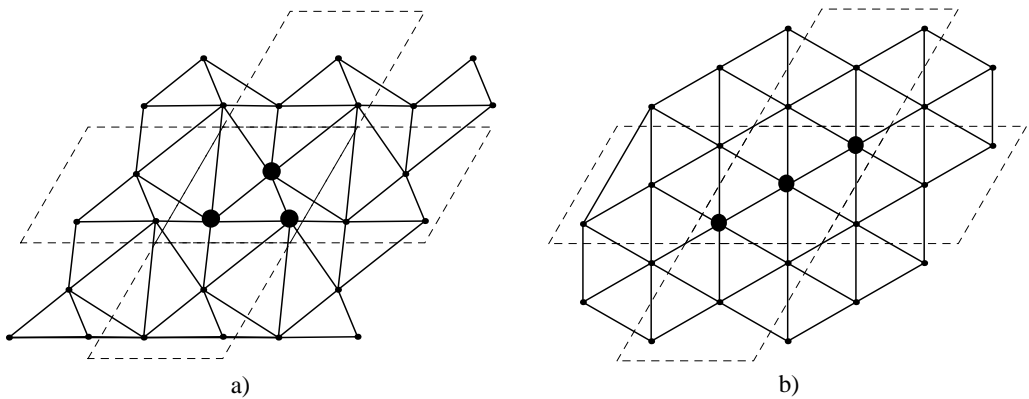} 
\caption{a) Three points in the cell $Q_0$ are distinguished. Dashed lines show the lattice, solid lines the double periodic Delaunay graph. b) The optimal graph isomorphic to the graph from a).} 
\label{fig:ex1}
\end{figure} 
Application of the algorithm yields the optimal hexagonal structure  Fig.\ref{fig:ex1}b. 
\end{example}

\begin{example}
\label{ex:2}
Consider the hexagonal lattice as in Example \ref{ex:1} and $N=16$ points with the corresponding double periodic Voronoi tessellation shown in Fig.\ref{fig:bio4}. The considered structure determines a double periodic graph $(\mathbf A,E)$. This graph generates the class of isomorphic graphs $\mathcal G_{(\mathbf A,E)}$. Find the optimal graph $(\mathbf A',E)$ in the class $\mathcal G_{(\mathbf A,E)}$. Construct the Voronoi tessellation corresponding to the set $\mathbf A'$ and the corresponding graph $(\mathbf A',E')$ which determines the new class $\mathcal G'_{(\mathbf A',E')}$. The optimal graph in the class $\mathcal G'_{(\mathbf A',E')}$ is the graph $(\mathbf A',E')$. Therefore, in this example the graph $(\mathbf A,E)$ from Fig.\ref{fig:bio4} is transformed into the graph $(\mathbf A',E')$ from Fig.\ref{fig:bioEx}.  
\begin{figure}[htp]
\centering
\includegraphics[clip, trim=0mm 0mm 0mm 0mm, width=0.8\textwidth]{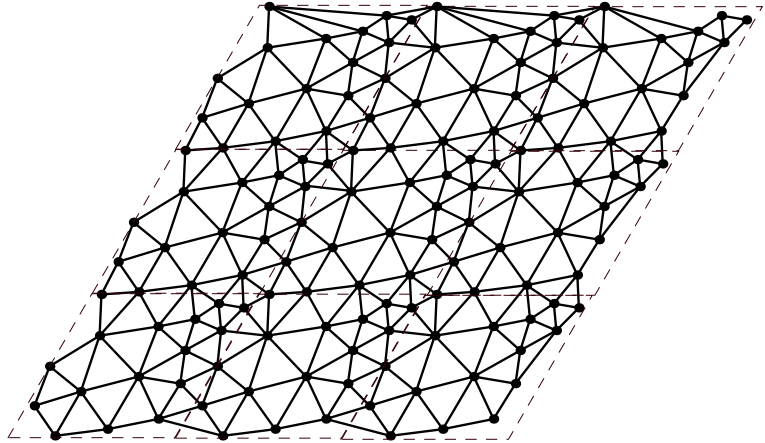} 
\caption{The optimal graph $(\mathbf A',E')$ from Example \ref{ex:2}.} 
\label{fig:bioEx}
\end{figure}
\end{example}

Every edge of the Delaunay graph models the interaction caused by the diffusion flux. This flux between two disks can be insignificant if the gap between the disk is sufficiently large. In this case, the corresponding edge should be deleted from the Delaunay graph. We consider such a case in the following example.      

\begin{example}
\label{ex:3}
Consider $N=9$ points and the double periodic Voronoi tessellation isomorphic to the hexagonal lattice. The perfect hexagonal array in Fig.\ref{fig:ex3}a presents the optimal graph $(\mathbf A,E)$. Consider another graph $(\mathbf A,E')$ obtained from $(\mathbf A,E)$ by deletion of the edges connecting the first layer of disks with other layers. In this case, the optimal graph becomes similar to hexagonal-layered structure displayed in Fig.\ref{fig:ex3}b. 
\begin{figure}[htp]
\centering
\includegraphics[clip, trim=0mm 0mm 0mm 0mm, width=1.0\textwidth]{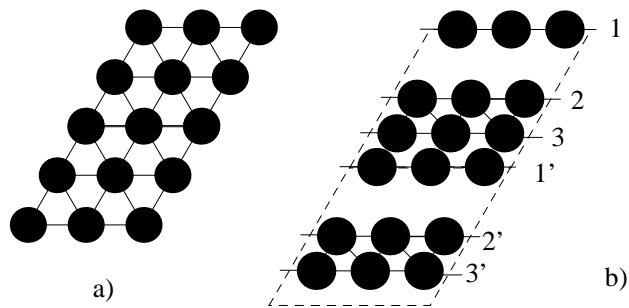} 
\caption{The optimal graphs from Example \ref{ex:3}. The disks in the layers 1 and 1',  2 and 2',  3 and 3' are the same in the toroidal topology.} 
\label{fig:ex3}
\end{figure}
\end{example}

The above examples illustrate few scenarios of the 2D pattern formations. Systematic simulations can help to study properties of the optimal graphs. The main feature of the proposed method is investigation of the graph structures by analytical and numerical methods without treatment of PDE. The method of structural approximation recalls a finite element method when a continuous problem is approximated by a discrete problem. 
The structural approximation is based on a "physical discretization" \cite{LB}, when edges of the graph correspond to the most intensive places of the diffusion flux. Further, the principle of minimum energy yields a discrete numerical problem as in a finite element method. The method of structural approximation was justified for the $p-$Laplacian including linear equations in \cite{LB}, \cite{Kolpakov}, \cite{ryl2008}.   

\begin{remark}
Solution to the optimal energy problem yields solution to the optimal packing problem for disks \cite{Mit2012}, \cite{Mit2014}. 
The corresponding concentration $\nu(\mathcal G)$ of disks attains the maximal value in the class $\mathcal G_{(\mathbf A,E)}$. The set of optimal graphs includes graphs corresponding to packing constructed by various packing protocols. This scheme gives the set of the optimal concentrations depending on protocols, i.e., on the class of graphs. Therefore, in order to get the set of all optimal packings, it is sufficient to determine the targets of the optimal locations \eqref{eq:sa25}. In the above examples, the graph chain \eqref{eq:sa25} terminates. We cannot give an example with an infinite  graph chain.  
\end{remark}

\end{document}